\documentclass[aps,pra,showpacs,amsmath,amssymb,amsfonts,tightenlines,superscriptaddress,twocolumn,lengthcheck]{revtex4-1}
\usepackage{amsmath}
\usepackage{amsfonts}
\usepackage{graphicx}
\usepackage{epsfig}
\usepackage{color}
\usepackage[colorlinks,citecolor=blue]{hyperref}

\begin{document}

\title{Simultaneous cooling of coupled mechanical resonators in cavity optomechanics}
\author{Deng-Gao Lai}
\affiliation{Key Laboratory of Low-Dimensional Quantum Structures and Quantum Control of Ministry of Education, Department of Physics and Synergetic Innovation Center for Quantum Effects and Applications, Hunan Normal University, Changsha 410081, China}
\author{Fen Zou}
\affiliation{Key Laboratory of Low-Dimensional Quantum Structures and Quantum Control of Ministry of Education, Department of Physics and Synergetic Innovation Center for Quantum Effects and Applications, Hunan Normal University, Changsha 410081, China}
\author{B. P. Hou}
\affiliation{College of Physics and Electronic Engineering, Institute of Solid State Physics, Sichuan Normal
University, Chengdu 610068, China}
\author{Yun-Feng Xiao}
\affiliation{State Key Laboratory for Mesoscopic Physics and School of Physics, Peking University; Collaborative Innovation Center of Quantum Matter, Beijing 100871, China}
\author{Jie-Qiao Liao}
\email{jqliao@hunnu.edu.cn}
\affiliation{Key Laboratory of Low-Dimensional Quantum Structures and Quantum Control of Ministry of Education, Department of Physics and Synergetic Innovation Center for Quantum Effects and Applications, Hunan Normal University, Changsha 410081, China}

\begin{abstract}
Quantum manipulation of coupled mechanical resonators has become an important research topic in optomechanics because these systems can be used to study the quantum coherence effects involving multiple mechanical modes. A prerequisite for observing macroscopic mechanical coherence is to cool the mechanical resonators to their ground state. Here we propose a theoretical scheme to cool two coupled mechanical resonators by introducing an optomechanical interface. The final mean phonon numbers in the two mechanical resonators are calculated exactly and the results show that the ground-state cooling is achievable in the resolved-sideband regime and under the optimal driving. By adiabatically eliminating the cavity field in the large-decay regime, we obtain analytical results of the cooling limits, which show the smallest achievable phonon numbers and the parameter conditions under which the optimal cooling is achieved. Finally, the scheme is extended to the cooling of a chain of coupled mechanical resonators.
\end{abstract}

\pacs{42.50.Pq, 42.50.Dv, 42.50.Wk, 42.50.Ct.}

\date{\today }
\maketitle

\section{Introduction}

The radiation-pressure coupling between electromagnetic fields and mechanical oscillation is at the heart of cavity optomechanics~\cite{Kippenberg2008Science,Meystre2013AP,Aspelmeyer2014RMP}. This coupling is the basis for both the control of the mechanical properties through the optical means and the manipulation of the field statistics by mechanically changing the cavity boundary~\cite{Rabl2011PRL,Nunnenkamp2011,Liao2012,Kronwald2013PRA,Liao2013,Liao2013PRA,Xu2013PRA0,Agarwal2010PRA,Hou2015PRA}.
In recent several years, much attention has been paid to optomechanical systems involving multiple mechanical resonators~\cite{Bhattacharya2008PRA,Xuereb2012PRL,Nair2016PRA,Stamper-Kurn2016NP,Massel2017PRA,Favero2017PRL}. This is because the multimode mechanical systems can be used to study macroscopic mechanical coherence such as quantum entanglement~\cite{Paz2008PRL,Xu2013PRA,Wang2013PRL,Liao2014PRA,Wang2016PRA} and quantum synchronization~\cite{Mari2013PRL,Matheny2014PRL}. Moreover, coupled mechanical systems have been widely applied to sensors for detecting various physical signals~\cite{Schwab2007NL,Roukes2012NL}, especially in nanomechanical systems~\cite{Schwab2007NL,Roukes2012NL,Yamaguchi2013NP,Huang2013PRL,Huang2016PRL}.

To observe the signature of quantum effects in mechanical systems, a prerequisite might be the cooling of the systems to theirs ground states such that the thermal noise can be suppressed. So far, several physical mechanisms such as feedback cooling~\cite{Mancini1998PRL,Cohadon1999PRL,Kleckner2006Nature,Corbitt2007PRL,Poggio2007PRL}, backaction cooling~\cite{Schliesser2006PRL,Peterson2016PRL}, and sideband cooling~\cite{Wilson-Rae2007PRL,Marquardt2007PRL,Xue2007PRB,Brown2007PRL,Genes2008PRA,Li2008PRB,Li2011PRA,Liu2017PRA,Xu2017PRL} have been proposed to cool a single mechanical resonator in optomechanics. Moreover, various new schemes have been proposed to cool mechanical resonators, such as transient cooling~\cite{Liao2011PRA,Machnes2012PRL}, cooling based on the quantum  interference effect~\cite{Xia2009PRL,Wang2011PRL,Li2011PRB,Yan2016PRA}, and quantum cooling in the strong-optomechanical-coupling regime~\cite{Liu2013PRL,Liu2014PRA}. In particular, the ground-state cooling has been realized in typical optomechanical systems, which is composed by a single cavity mode and a single mechanical mode~\cite{Chan2011Nature,Teufel2011Nature}. Correspondingly, to manipulate the quantum coherence in multimode optomechanical systems, it is desired to cool these mechanical modes for further quantum manipulation. Nevertheless, how to cool multiple mechanical resonators remains an unclear question.

In this paper, we present a practical scheme to cool two coupled mechanical resonators in an optomechanical system which is formed by an optomechanical cavity coupled to another mechanical resonator. Here the two mechanical resonators are coupled to each other through the so-called ``position-position" coupling. In the strong-driving regime, the system is linearized to a three-mode cascade system, which is composed by a cavity mode and two mechanical modes. To include the cooling channel and the environments, we assume that the cavity field is connected with a vacuum bath, and the two mechanical resonators are connected with two heat baths at finite temperatures. Physically, the vacuum bath of the cavity field extracts the thermal excitations in the mechanical resonators via a manner of nonequilibrium dynamics, and then the total system reaches a steady state. By exactly calculating the final mean phonon numbers in the resonators, we find that the ground-state cooling of the two mechanical resonators can be realized simultaneously under the optimal driving detuning and in the resolved-sideband regime. Specifically, the cooling limits of the two mechanical resonators are analytically derived by adiabatically eliminating the cavity field in the large-decay regime. Finally, we extend the optomechanical scheme to the cooling of a chain of coupled mechanical resonators. The results show that ground state cooling is achievable in multiple mechanical resonators, and that the cooling efficiency is higher for the mechanical oscillator, which is closer to the cavity.

The rest of this paper is organized as follows. In Sec.~\ref{sec2}, we introduce the physical model and present the Hamiltonians. In Sec.~\ref{sec3}, we derive the equations of motion and find the solutions. In Sec.~\ref{sec4}, we calculate the final mean phonon numbers, analyze the parameter dependence, and derive the cooling limits of the two mechanical resonators. In Sec.~\ref{sec5}, we extend our studies to the case of a chain of coupled mechanical resonators. Finally, we present a brief conclusion in Sec.~\ref{sec6}. Two appendixes are presented to display the detailed calculations of the final mean phonon numbers and the cooling limits.

\section{Model and Hamiltonian\label{sec2}}

\begin{figure}[tbp]
\center
\includegraphics[bb=42 551 260 749, width=0.47 \textwidth]{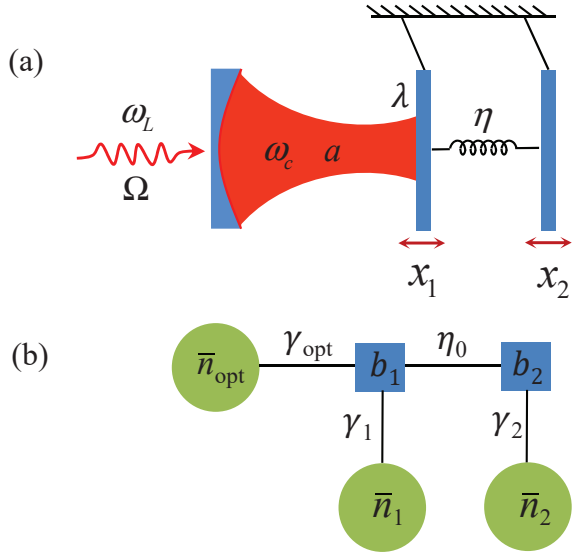}
\caption{(a) Schematic of the three-mode optomechanical system. A single-mode cavity field with resonance frequency $\omega_{c}$ is coupled to an oscillating end mirror with resonance frequency $\tilde{\omega}_{1}$ via the radiation-pressure coupling. The movable end mirror is coupled to another mechanical resonator with resonance frequency $\tilde{\omega}_{2}$ via the ``position-position" interaction. (b) By adiabatically eliminating the cavity mode, the model in panel (a) is simplified to a system of two coupled mechanical modes $b_{1}$ and $b_{2}$, with the coupling strength $\eta_{0}$. The mechanical resonator $b_{l=1,2}$ is coupled to the heat bath with the decay rate $\gamma_{l}$ and thermal occupation number $\bar{n}_{l}$. Additionally, the mode $b_{1}$ is coupled to an effective optical bath with the effective decay rate $\gamma_{\text{opt}}$ and thermal occupation number $\bar{n}_{\text{opt}}$.}
\label{Figmodel}
\end{figure}

We consider a three-mode optomechanical system, which is composed of one cavity mode and two mechanical modes, as illustrated in Fig.~\ref{Figmodel}(a). The cavity field mode is coupled to the first mechanical mode via the radiation-pressure coupling, and the two mechanical modes are coupled to each other via the so-called ``position-position" interaction. To manipulate the optical and mechanical degrees of freedom, a proper driving field is applied to the optical cavity.
The Hamiltonian of the system reads ($\hbar =1$)
\begin{eqnarray}
H &=&\omega _{c}a^{\dagger }a+\sum_{l=1,2}\left(\frac{p_{xl}^{2}}{2m_{l}}+\frac{m_{l}\tilde{\omega}_{l}^{2}x_{l}^{2}}{2}\right)-\lambda a^{\dagger }ax_{1}\notag \\
&&+\eta \left( x_{1}-x_{2}\right) ^{2}+\Omega(a^{\dagger }e^{-i\omega_{L}t}+ae^{i\omega_{L}t}),\label{eq1iniH}
\end{eqnarray}
where $a$ and $a^{\dagger }$ are, respectively, the annihilation and creation operators
of the cavity mode with the resonance frequency $\omega_{c}$. The coordinate and momentum operators $x_{l}$ and $p_{xl}$ are introduced to describe the $l$th ($l=1,2$) mechanical resonator with mass $m_{l}$ and resonance frequency $\tilde{\omega}_{l}$. The optomechanical coupling between the cavity field and the first mechanical mode is described by the $\lambda$ term in Eq.~(\ref{eq1iniH}), where $\lambda=\omega_{c}/L$ denotes the optomechanical force of a single photon, with $L$ being the rest length of the optical cavity.
The $\eta$ term depicts the mechanical interaction between the two mechanical resonators. The parameters $\omega_{L}$  and $\Omega$ are, respectively, the optical driving frequency and driving amplitude, which is determined by the driving power via the relation $\Omega=\sqrt{2P_{L}\kappa/\omega_{L}}$, where $P_{L}$ is the power of the driving laser, and $\kappa$ is the decay rate of the cavity field.

For below convenience, we introduce the normalized resonance frequencies $\omega_{l=1,2}=\sqrt{\tilde{\omega}_{l}^{2}+2\eta/m_{l}}$ and the dimensionless position and momentum operators $q_{l=1,2}=\sqrt{m_{l}\omega_{l}}x_{l}$ and $p_{l=1,2}=\sqrt{1/(m_{l}\omega_{l})}p_{xl}$ ($[q_{l},p_{l}]=i$) for the mechanical resonators. Then, in the rotating frame defined by the unitary transformation operator $\exp(-i\omega_{L}ta^{\dagger}a)$, the Hamiltonian of the system becomes
\begin{eqnarray}
H_{I}&=&\Delta_{c}a^{\dagger}a+\sum_{l=1,2}\frac{\omega_{l}}{2}(q_{l}^{2}+p_{l}^{2})-\lambda_{0}a^{\dagger}aq_{1}\nonumber \\
&&-2\eta_{0}q_{1}q_{2}+\Omega(a^{\dagger}+a),\label{Hamlt2dimless}
\end{eqnarray}
where $\Delta_{c}=\omega_{c}-\omega_{L}$ is the driving detuning of the cavity field, $\lambda_{0}=\lambda\sqrt{1/(m_{1}\omega_{1})}$ and $\eta_{0}=\eta \sqrt{1/(m_{1}m_{2}\omega_{1}\omega_{2})}$ denote the strength of the optomechanical coupling and the mechanical coupling in the dimensionless representation, respectively. The Hamiltonian~(\ref{Hamlt2dimless}) is the starting point of our consideration.  Below, we will study the cooling performance by seeking the steady-state solution of the system.

\section{The Langevin equations\label{sec3}}

Quantum systems are inevitably coupled to their environments. To treat the damping and noise in our model, we consider the case where the optical mode is linearly coupled to a vacuum bath and the two mechanical modes experience the Brownian motion. In this case, the evolution of the system can be described by the Langevin equations,
\begin{subequations}
\label{Langevineqorig}
\begin{align}
\dot{a}&=-[\kappa+i(\Delta_{c}-\lambda_{0}q_{1})]a-i\Omega+\sqrt{2\kappa}a_{\textrm{in}} , \\
\dot{q}_{l}&=\omega_{l}p_{l}, \hspace{1 cm}l=1,2, \\
\dot{p}_{1}&=-\omega_{1}q_{1}-\gamma_{1}p_{1}+\lambda_{0}a^{\dagger}a+2\eta_{0}q_{2}+\xi_{1}, \\
\dot{p}_{2}&=-\omega_{2}q_{2}-\gamma_{2}p_{2}+2\eta_{0}q_{1}+\xi_{2},
\end{align}
\end{subequations}
where $\kappa$ and $\gamma_{l=1,2}$ are the decay rates of the cavity mode and the $l$th mechanical mode, respectively. The operators $a_{\textrm{in}}$ $(a^{\dagger}_{\textrm{in}})$ and $\xi_{l=1,2}$ are the noise operator of the cavity field and the Brownian force act on the $l$th mechanical resonator, respectively. These operators have zero mean values and the following correlation functions,
\begin{subequations}
\label{correlationfun}
\begin{align}
\langle a_{\textrm{in}}(t) a_{\textrm{in}}^{\dagger}(t^{\prime})\rangle=&\delta(t-t^{\prime}), \hspace{0.5 cm}
\langle a_{\textrm{in}}^{\dagger}(t) a_{\textrm{in}}(t^{\prime})\rangle =0, \\
\langle \xi_{l}(t)\xi_{l}(t^{\prime})\rangle=&\frac{\gamma _{l}}{\omega _{l}}\int \frac{d\omega }{2\pi }e^{-i\omega(t-t^{\prime}) }\omega  \notag \\
&\times\left[\coth\left(\frac{\omega}{2k_{B}T_{l}}\right) +1\right],
\end{align}
\end{subequations}
where $k_{B}$ is the Boltzmann constant, and $T_{l=1,2}$ is the bath temperature of the $l$th mechanical resonator.

To cool the mechanical resonators, we consider the strong-driving regime of the cavity such that the average photon number in the cavity is sufficient large and then the linearization procedure can be used to simplify the physical model. To this end, we express the operators in Eq.~(\ref{Langevineqorig}) as the sum of their steady-state mean values and quantum fluctuations, namely $o=\left\langle o\right\rangle_{\textrm{ss}} +\delta o$
for operators $a$, $a^{\dagger}$, $q_{l=1,2}$, and $p_{l=1,2}$. By separating the classical motion and the quantum fluctuation, the linearized equations of motion for the quantum fluctuations can be written as
\begin{subequations}
\label{linzedlandeveq}
\begin{align}
\delta\dot{a}=&-(\kappa+i\Delta)\delta a+iG\delta q_{1}+\sqrt{2\kappa}a_{\textrm{in}},\\
\delta\dot{q}_{l}=&\omega_{l}\delta p_{l}, \hspace{1 cm}l=1,2,\\
\delta\dot{p}_{1}=&-\omega_{1}\delta q_{1}-\gamma_{1}\delta p_{1}+2\eta_{0}\delta q_{2}\notag\\
&+G^{\star}\delta a+G\delta a^{\dagger}+\xi_{1},\\
\delta\dot{p}_{2}=&-\omega_{2}\delta q_{2}-\gamma_{2}\delta p_{2}+2\eta_{0}\delta q_{1}+\xi_{2},
\end{align}
\end{subequations}
where $\Delta=\Delta_{c}-\lambda_{0}\langle q_{1}\rangle_{\textrm{ss}}$ is the driving detuning normalized by the linearization and $G=\lambda_{0}\langle a\rangle_{\textrm{ss}}$ denotes the strength of the linearized optomechanical coupling. Here, the steady-state solution of the quantum Langevin equations in Eq.~(\ref{Langevineqorig}) can be obtained as
\begin{subequations}
\begin{align}
\langle a\rangle_{\textrm{ss}}=&\frac{-i\Omega}{\kappa+i\Delta}, \\
\langle q_{1}\rangle_{\textrm{ss}}=&\frac{\lambda_{0}\omega_{2}\langle a^{\dagger}\rangle_{\textrm{ss}}\langle a\rangle_{\textrm{ss}}}{\omega_{1}\omega_{2}-4\eta_{0}^{2}},\\
\langle q_{2}\rangle_{\textrm{ss}}=&\frac{2\lambda_{0}\eta_{0}\langle a^{\dagger}\rangle_{\textrm{ss}}\langle a\rangle_{\textrm{ss}}}{\omega_{1}\omega_{2}-4\eta_{0}^{2}},\\
\langle p_{1}\rangle_{\textrm{ss}}=&\langle p_{2}\rangle_{\textrm{ss}}=0.
\end{align}
\end{subequations}

The cooling problem can be solved by calculating the steady-state solution of Eq.~(\ref{linzedlandeveq}). This can be realized by solving the variables in the frequency domain with the Fourier transformation method. Under the definition for operator $r$ ($r=\delta a$, $\delta q_{l}$, $\delta p_{l}$, $a_{in}$, $\xi$) and its conjugate $r^{\dagger}$,
\begin{subequations}
\begin{eqnarray}
r(t) &=&\frac{1}{\sqrt{2\pi }}\int_{-\infty }^{\infty }e^{i\omega t}\tilde{r}(\omega) d\omega,\\
r^{\dagger }(t) &=&\frac{1}{\sqrt{2\pi }}\int_{-\infty }^{\infty}e^{-i\omega t}\tilde{r}^{\dagger }(\omega) d\omega,
\end{eqnarray}
\end{subequations}
the equations of motion~(\ref{linzedlandeveq}) can be expressed in the frequency domain as
\begin{subequations}
\label{fluceqinfdomain}
\begin{align}
i\omega\delta\tilde{a}(\omega)=&-(\kappa+i\Delta)\delta\tilde{a}(\omega)+iG\delta\tilde{q}_{1}(\omega)+\sqrt{2\kappa }\tilde{a}_{in}(\omega),\\
i\omega\delta\tilde{q}_{l}(\omega)=&\omega_{l}\delta\tilde{p}_{l}(\omega),  \hspace{1 cm}l=1,2,\\
i\omega\delta\tilde{p}_{1}(\omega)=&-\omega_{1}\delta\tilde{q}_{1}(\omega)-\gamma_{1}\delta\tilde{p}_{1}(\omega)+G^{\star}\delta\tilde{a}(\omega)  \notag \\
&+G\delta\tilde{a}^{\dagger}(\omega)+2\eta_{0}\delta\tilde{q}_{2}(\omega)+\tilde{\xi}_{1}(\omega),\\
i\omega\delta\tilde{p}_{2}(\omega)=&-\omega_{2}\delta\tilde{q}_{2}(\omega)-\gamma_{2}\delta\tilde{p}_{2}(\omega)   +2\eta_{0}\delta \tilde{q}_{1}(\omega)\notag \\
&+\tilde{\xi}_{2}(\omega),
\end{align}
\end{subequations}
which can be further solved as
\begin{subequations}
\label{fdomainsolution}
\begin{align}
\delta \tilde{a}(\omega)=&\{iGC_{1}(\omega)\tilde{a}_{\text{in}}^{\dagger}(\omega)+[iGC_{1}^{\star }(-\omega)+\sqrt{
2\kappa}B(\omega)]\tilde{a}_{\text{in}}(\omega)   \nonumber \\
&+iGW_{1}(\omega)\tilde{\xi}_{1}(\omega)+iGW_{2}(\omega)
\tilde{\xi}_{2}(\omega)\}\nonumber \\
&\times [ \kappa +i( \Delta +\omega)]^{-1}B^{-1}(\omega ),\\
\delta\tilde{q}_{1}(\omega)=&[C_{1}(\omega )\tilde{a}_{\text{in}}^{\dagger}(\omega)+C_{1}^{\star }(-\omega )\tilde{a}_{\text{in}}(\omega) +W_{1}(\omega )\tilde{\xi}_{1}(\omega) \nonumber \\
&+W_{2}(\omega )\tilde{\xi}_{2}(\omega)]B^{-1}(\omega ),\\
\delta \tilde{q}_{2}(\omega)=&[C_{2}(\omega )\tilde{a}_{\text{in}}^{\dagger}(\omega)+C_{2}^{\star }(-\omega )\tilde{a}_{\text{in}}(\omega)+W_{2}(\omega )\tilde{\xi}_{1}(\omega )   \nonumber \\
&+W_{3}(\omega )\tilde{\xi}_{2}(\omega)]B^{-1}(\omega ),
\end{align}
\end{subequations}
where we introduced the variables
\begin{subequations}
\begin{align}
B(\omega)=&(i\gamma_{1}\omega-\omega^{2}+\omega_{1}^{2})(-i\gamma_{2}\omega+\omega^{2}-\omega_{2}^{2})[(\kappa+i\omega)^{2}\nonumber \\
&+\Delta^{2}]+2\omega_{1}(i\gamma_{2}\omega-\omega^{2}+\omega_{2}^{2})|G|^{2}\Delta\nonumber\\
&+4\omega_{1}\omega_{2}\eta_{0}^{2}[(\kappa+i\omega)^{2}+\Delta^{2}],\\
C_{1}(\omega)=&\sqrt{2\kappa}G\omega_{1}[\gamma_{2}\omega+i(\omega^{2}-\omega_{2}^{2})](-i\kappa+\omega+\Delta),\\
C_{2}(\omega)=&-2\sqrt{2\kappa}\eta_{0}G\omega_{1}\omega_{2}[\kappa+i(\omega+\Delta)],\\
W_{1}(\omega)=&\omega_{1}(-i\gamma_{2}\omega+\omega^{2}-\omega_{2}^{2})[(\kappa+i\omega)^{2}+\Delta^{2}],\\
W_{2}(\omega)=&-2\eta_{0}\omega_{1}\omega_{2}[(\kappa+i\omega)^{2}+\Delta^{2}],\\
W_{3}(\omega)=&2\omega_{1}\omega_{2}\vert G\vert^{2}\Delta+\omega_{2}(-i\gamma_{1}\omega+\omega^{2}-\omega_{1}^{2})\nonumber\\
&\times[(\kappa+i\omega)^{2}+\Delta^{2}].
\end{align}
\end{subequations}
In principle, the expressions of these quantum fluctuations $\delta a$, $\delta q_{l=1,2}$, and $\delta p_{l=1,2}$ in the time domain can be calculated by performing the inverse Fourier transformation. For our cooling task, we will focus on the steady-state mean value of the phonon numbers in the mechanical resonators.

In the above consideration, we do the linearization around the steady state of the system. Therefore, we need to analyze the stability of the system. By applying the Routh-Hurwitz criterion~\cite{Gradstein2014book}, it is found that the stability condition, under which the system reaches a steady state, is given by
\begin{eqnarray}
\Delta_{6}>0,
\end{eqnarray}
where the expression of $\Delta_{6}$ is defined in Eq.~(\ref{Delta6def}). In the following consideration, all the used parameters satisfy this stability condition.

\section{Cooling of two mechanical resonators\label{sec4}}

In this section, we study the cooling efficiency of the mechanical resonators by calculating the final mean phonon numbers and deriving the cooling limits.

\subsection{The final mean phonon numbers\label{secabpn}}

For the purpose of quantum cooling, we prefer to calculate the fluctuation spectrum of the position and momentum operators for the two mechanical resonators, and then the final mean phonon numbers in the mechanical resonators can be obtained by integrating the corresponding fluctuation spectra. Mathematically, the final mean phonon numbers in the two mechanical resonators can be calculated by the relation~\cite{Genes2008PRA}
\begin{equation}
n^{f}_{l}=\frac{1}{2}[\langle\delta q_{l}^{2}\rangle +\langle\delta p_{l}^{2}\rangle-1],\label{finalphonumber}
\end{equation}
where the variances $\delta q_{l}^{2}$ and $\delta p_{l}^{2}$ of the position and momentum operators can be obtained by solving Eq.~(\ref{fluceqinfdomain}) in the frequency domain and integrating the corresponding fluctuation spectra,
\begin{subequations}
\label{specintegral}
\begin{align}
\langle\delta q_{l}^{2}\rangle=&\frac{1}{2\pi}\int_{-\infty}^{\infty}S_{q_{l}}(\omega)d\omega,\hspace{1 cm}l=1,2,\\
\langle\delta p_{l}^{2}\rangle=&\frac{1}{2\pi}\int_{-\infty}^{\infty}S_{p_{l}}(\omega)d\omega\nonumber\\
=&\frac{1}{2\pi\omega^{2}_{l}}\int_{-\infty}^{\infty}\omega^{2}S_{q_{l}}(\omega)d\omega,\hspace{1 cm}l=1,2.
\end{align}
\end{subequations}
Here the fluctuation spectra of the position and momentum of the two mechanical oscillators are defined by
\begin{equation}
S_{o}(\omega)=\int_{-\infty}^{\infty}e^{-i\omega\tau}\langle \delta o(t+\tau) \delta o(t)\rangle_{\textrm{ss}}d\tau,\label{spectrumtimedomain}
\end{equation}
for $o=q_{l=1,2}$ and $p_{l=1,2}$. Here the average $\langle \cdot\rangle_{\textrm{ss}}$ are taken over the steady state of the system.
The fluctuation spectrum can also be expressed in the frequency domain as
\begin{equation}
\langle\delta\tilde{o}(\omega)\delta\tilde{o}(\omega')\rangle_{\textrm{ss}}=S_{o}(\omega) \delta(\omega+\omega'), \hspace{0.5 cm}(o=q_{l},p_{l}).\label{spectrumfdomain}
\end{equation}
Based on the results given in Eqs.~(\ref{fdomainsolution}), (\ref{spectrumfdomain}), and the correlation function (\ref{correlationfun}) in the frequency domain, the position and momentum fluctuation spectra of the mechanical resonators can be obtained as
\begin{subequations}
\label{spectraexp}
\begin{align}
S_{q_{1}}\left( \omega \right)  =&\frac{1}{\left\vert B(\omega )\right\vert^{2}}\left\{\left\vert C_{1}(\omega )\right\vert ^{2} \right. \nonumber \\
&\left.+\left\vert W_{1}(\omega)\right\vert ^{2}\frac{\gamma _{1}\omega }{\omega _{1}}\left[1+\coth\left( \frac{\omega }{2k_{\textrm{B}}T_{1}}\right)\right]\right. \nonumber \\
&\left.+\left\vert W_{2}(\omega)\right\vert ^{2}  \frac{\gamma _{2}\omega }{\omega _{2}}\left[1+\coth\left( \frac{\omega }{2k_{\textrm{B}}T_{2}}\right)\right]\right\},\\
S_{q_{2}}\left( \omega \right)  =&\frac{1}{\left\vert B(\omega )\right\vert^{2}}\left\{\left\vert C_{2}(\omega )\right\vert ^{2}\right.  \nonumber \\
&\left.+\left\vert W_{2}(\omega)\right\vert ^{2}\frac{\gamma _{1}\omega }{\omega _{1}}\left[1+\coth\left( \frac{\omega }{2k_{\textrm{B}}T_{1}}\right)\right]\right.  \nonumber \\
&\left.+\left\vert W_{3}(\omega)\right\vert ^{2}\frac{\gamma _{2}\omega }{\omega _{2}}\left[1+\coth\left( \frac{\omega }{2k_{\textrm{B}}T_{2}}\right)\right]\right\},\\
S_{p_{l}}(\omega )=&\left(\frac{\omega }{\omega _{l}}\right)^{2}S_{q_{l}}(\omega), \hspace{0.5 cm}l=1,2.
\end{align}
\end{subequations}
In terms of Eqs.~(\ref{finalphonumber}), (\ref{specintegral}), and (\ref{spectraexp}), the exact analytical results of the final mean phonon numbers in the two mechanical resonators can be obtained (see Appendix~\ref{appendixa} for details).

\subsection{Ground state cooling}

Based on the above results, we now study the cooling of the two coupled mechanical resonators by the optomechanical coupling. Physically, the system becomes, by linearization, a chain of three modes with the bilinear-type coupling between the neighboring modes. As a result, the excitation energy can be exchanged between the two neighboring modes by the rotating-wave term (namely the beam-splitter-type coupling) in the near-resonance and weak-coupling regimes. In this system, the two mechanical resonators are connected to two heat baths and the cavity field is connected to a vacuum bath. Hence the final mean phonon numbers in the two mechanical resonators would be finite numbers, which should be smaller than the thermal phonon occupations in the heat baths because the thermal excitations can be finally extracted to the vacuum bath. In this sense, the mechanical resonators can be cooled by the optomechanical coupling. Below, we will show how do the final mean phonon numbers in the two mechanical resonators depend on the parameters of the system.
\begin{figure}[tbp]
\centering
\includegraphics[bb=100 217 462 702, width=0.47 \textwidth]{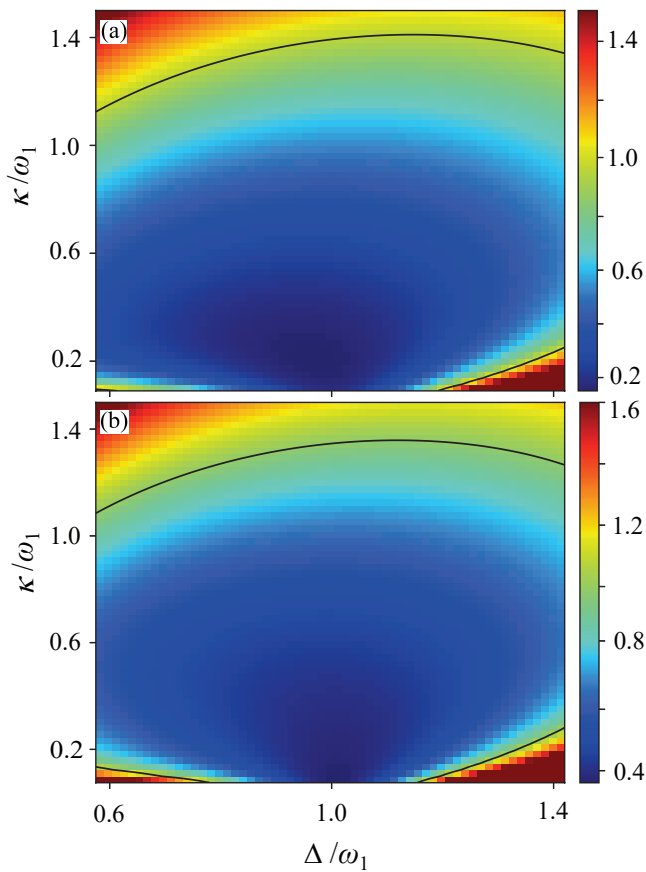}
\caption{(Color online) The final mean phonon numbers (a) $n^{f}_{1}$ and (b) $n^{f}_{2}$ in the two mechanical resonators vs the effective driving detuning $\Delta/\omega_{1}$ and the decay rate $\kappa/\omega_{1}$. The used parameters are given by $\omega_{1}/2\pi=\omega_{2}/2\pi=10$ MHz, $\gamma_{1}/\omega_{1}=\gamma_{2}/\omega_{1}=10^{-5}$, $\omega_{c}/\omega_{1}=2.817\times10^{7}$, $\eta_{0}/\omega_{1}=0.04$, $m_{1}=m_{2}=250$ ng, $\bar{n}_{1}=\bar{n}_{2}=1000$, $L=0.5$ mm, $P_{L}=50$ mW, and $\lambda=1064$ nm. The black solid curves correspond to $n^{f}_{1}=n^{f}_{2}=1$.}\label{detuningdecay}
\end{figure}

In Fig.~\ref{detuningdecay}, we plot the final mean phonon numbers $n_{1}^{f}$ and $n_{2}^{f}$ as a function of the driving detuning $\Delta/\omega_{1}$ and the cavity-field decay rate $\kappa/\omega_{1}$. Here we choose the mechanical frequency $\omega_{1}$ as the frequency scale so that we can clearly see the relationship between the optimal driving detuning and the phonon sidebands, and the influence of the sideband-resolution condition on the cooling performance. When $\kappa/\omega_{1}\ll1$, the phonon sidebands can be resolved from the cavity emission spectrum, and this regime is called as the resolved-sideband limit. We can see from Fig.~\ref{detuningdecay} that the two resonators can be cooled  efficiently ($n^{f}_{1},n^{f}_{2}\ll1$) in the resolved-sideband limit and under the driving $\Delta/\omega_{1}\sim1$, which means that the ground-state cooling is achievable in this system. For the used parameters, the minimum phonon numbers for the two resonators are $n^{f}_{1}\approx0.15$ and $n^{f}_{2}\approx0.35$. For a given value of the ratio $\kappa/\omega_{1}$, the optimal driving detuning is given by $\Delta\approx\omega_{1}$. This is because the energy extraction efficiency between the cavity mode and the first mechanical mode should be maximum at $\Delta=\omega_{1}$, and the small deviation of the exact value of $\omega_{1}$ in realistic simulations is caused by the counter rotating-wave term in the linearized interaction between the cavity mode and the first mechanical mode. Physically, the generation of an anti-Stokes photon will cool the mechanical oscillator by taking away a phonon from the mechanical resonator. For the optimal cooling detuning $\Delta\approx\omega_{1}$, the frequency $\omega_{1}$ of the phonon exactly matches the driving detuning $\Delta$ and hence $\Delta/\omega_{1}=1$ corresponds to the optimal cooling. At the optimal driving $\Delta=\omega_{1}$, the final mean phonon numbers become worse with the increase of the ratio $\kappa/\omega_{1}$. In order to clearly illustrate the dependence of the mean phonon numbers on the parameters, we show a rough boundary of ground state cooling ($n^{f}_{1}$ and $n^{f}_{2}=1$), as shown by the black solid curves in Fig.~\ref{detuningdecay}(a) and Fig.~\ref{detuningdecay}(b). These results are consistent with the sideband cooling results in a typical optomechanical system~\cite{Marquardt2007PRL,Wilson-Rae2007PRL,Genes2008PRA,Li2008PRB}.

\begin{figure}[tbp]
\centering
\includegraphics[bb=52 13 342 286, width=0.47 \textwidth]{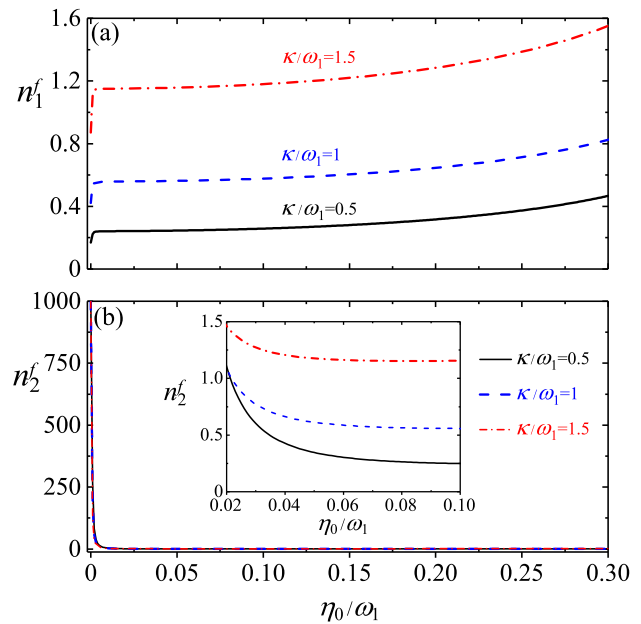}
\caption{(Color online) The final mean phonon numbers (a) $n^{f}_{1}$ and (b) $n^{f}_{2}$ as a function of $\eta_{0}/\omega_{1}$ when the cavity-field decay rate takes different values $\kappa/\omega_{1}=0.5$, $1$, and $1.5$. The inset in panel (b) is a zoomed-in plot of $n^{f}_{2}$ as a function of $\eta_{0}/\omega_{1}$, which shows clearly the dependence of $n^{f}_{2}$ on the cavity-field decay rate. Here we consider the optimal driving case $\Delta=\omega_{1}$, and other parameters are the same as those used in Fig.~\ref{detuningdecay}.}\label{couplingoftwo}
\end{figure}
Since the cavity provides the direct channel to extract the thermal excitations in the first mechanical resonator, the optimal driving (corresponding to a resonant beam-splitter-type interaction) is important to the cooling efficiency. At the same time, the coupling between the two mechanical resonators provides the channel to extract the thermal excitations from the second mechanical resonator, as a cascade cooling process. Consequently, the cooling efficiency of the second mechanical resonator should depend on the rotating-wave coupling between the two mechanical resonators, which is determined by the resonance frequencies of the two resonators and the coupling strength between them. To see this effect, in Fig.~\ref{couplingoftwo} we plot the final mean phonon numbers $n_{1}^{f}$ and $n_{2}^{f}$ as a function of the mechanical coupling strength $\eta_{0}$ between the two resonators when the cavity decay rate takes different values. Based on the fact that the second mechanical resonator will not be cooled at $\eta_{0}=0$, we confirm that the coupling between the two mechanical resonators provides the cooling channel for the second resonator. With the increase of $\eta_{0}$, the phonon number $n^{f}_{1}$ increases, while the phonon number $n^{f}_{2}$ decreases. This is because the first resonator provides the cool channel of the second resonator by extracting its thermal excitations, while the second resonator will encumber the cooling efficiency of the first resonator. Additionally, the final mean phonon numbers are larger for larger values of the decay rate $\kappa/\omega_{1}$, which is consistent with the analyses concerning the dependence of the cooling efficiency on the sideband-resolution condition.

\begin{figure}[tbp]
\centering
\includegraphics[bb=33 12 345 287, width=0.47 \textwidth]{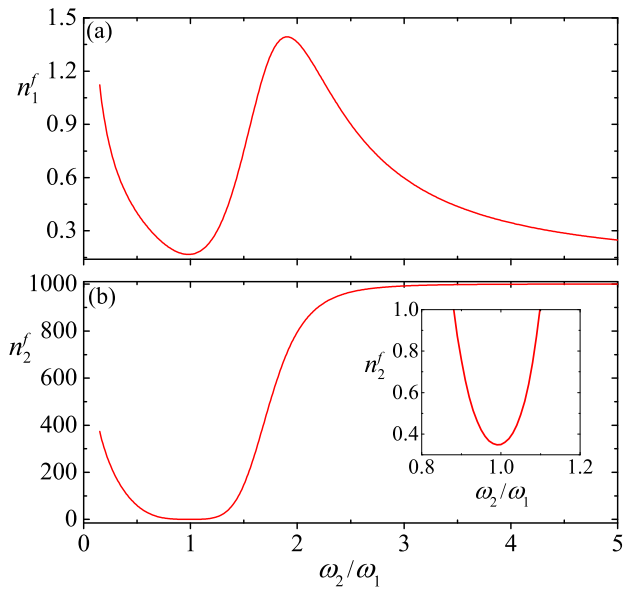}
\caption{(Color online) The final mean phonon numbers (a) $n^{f}_{1}$ and (b) $n^{f}_{2}$ vs the ratio $\omega_{2}/\omega_{1}$. The inset in panel (b) is a zoomed-in plot of $n^{f}_{2}$ as a function of $\omega_{2}/\omega_{1}$ from $0.8$ to $1.2$. Here we choose $\Delta=\omega_{1}$ and $\kappa/\omega_{1}=0.2$. Other parameters are the same as those given in Fig.~\ref{detuningdecay}.}\label{frequencyradio1}
\end{figure}
This cascade cooling process can also be seen by considering the case where the two mechanical resonators have different resonance frequencies. In Fig.~\ref{frequencyradio1}, we display the dependence of the final mean phonon numbers $n^{f}_{1}$ and $n^{f}_{2}$ on the frequency $\omega_{2}$ of the second resonator. Here we choose $\Delta=\omega_{\textrm{1}}$ such that the cooling efficiency of the first resonator is optimal. The result shows that both the two resonators have good cooling efficiency when the two resonators are resonant and near resonant ($\omega_{2}$ around $\omega_{1}$). With the increase of the detuning between the two resonance frequencies, the cooling efficiency becomes worse. The reason for this phenomenon is that the efficiency of energy extraction from the second resonator decreases with the increase of the detuning $\vert\omega_{1}-\omega_{2}\vert$, and that the counter-rotating-wave interaction terms, which simultaneously create phonon excitations in the two resonators, become important when the frequency detuning becomes comparable to the mechanical frequencies. When $\omega_{2}/\omega_{1}>2$, the cooling of the second resonator is almost turned off because the interaction between the two resonators is approximately negligible under the condition $\eta_{0}/\vert\omega_{1}-\omega_{2}\vert\ll1$. In this case, the cooling of the first resonator becomes better because the thermalization effect induced by the bath of the second resonator is turned off, and then the system is reduced to a typical optomechanical system with one cavity mode and one mechanical mode.

\begin{figure}[tbp]
\centering
\includegraphics[bb=27 17 400 260, width=0.47 \textwidth]{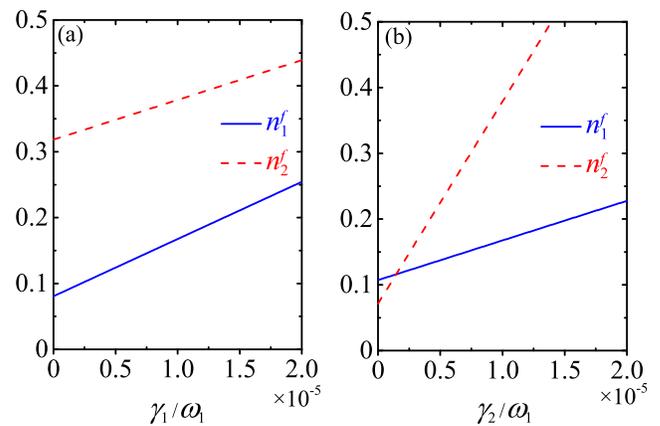}
\caption{(Color online) The final mean phonon numbers $n^{f}_{1}$ and $n^{f}_{2}$ as a function of (a) $\gamma_{1}$ and (b) $\gamma_{2}$. Here we take $\Delta=\omega_{1}$ and $\kappa/\omega_{1}=0.2$. Other parameters are the same as those given in Fig.~\ref{detuningdecay}.}\label{decay1decay2}
\end{figure}
We note that the final mean phonon numbers $n^{f}_{1}$ and $n^{f}_{2}$ in the two resonators also depend on the mechanical decay rates $\gamma_{1}$ and $\gamma_{2}$. In Fig.~\ref{decay1decay2} we show the final phonon numbers as a function of the decay rates. Here we see that $n^{f}_{1}$ and $n^{f}_{2}$ increase with the increase of the mechanical decay rates. This is because the energy exchange rates between the mechanical resonators and their heat baths are faster for larger values of the decay rates, and then the thermal excitation in the heat bath will raise the phonon numbers in the mechanical resonators.

In the plots in this section, we see that the first mechanical resonator is cooled better than the second resonator, i.e., $n^{f}_{1}<n^{f}_{2}$ under the same parameters.
This phenomenon is a physical consequence of the cascade cooling process in this system. The vacuum bath of the cavity plays the role of the pool to absorb the thermal excitations extracted from the two mechanical resonators. The cavity extracts the thermal excitations from the first resonator and transfers the excitations to its vacuum bath. The first resonator extracts the thermal excitations from the second resonator. Each mechanical resonator is connected to an independent heat bath, and the two heat baths have the same temperature. Hence, the relation $n^{f}_{1}<n^{f}_{2}$ can be understood from the point of view of the nonequilibrium physical process.

\subsection{The cooling limits}

Our exact results show that the ground-state cooling (with $n^{f}_{1,2}\ll1$) is achievable for the two mechanical resonators under proper parameters. However, what are the cooling limits (i.e., the smallest achievable phonon numbers) of the resonators remains unclear. In this section, we derive the approximate cooling results in the bad-cavity regime such that analytical expressions of the cooling limits can be obtained. This is achieved by eliminating adiabatically the cavity field in the large-decay regime ($\kappa\gg\tilde{G}$) and then calculating the final phonon numbers in the two mechanical modes under the rotating-wave approximation ($\omega_{1,2}\gg\tilde{G}$). In this case, the system is reduced to two coupled modes $b_{1}$ and $b_{2}$, where the mode $b_{1}$ is contacted with the optomechanical cooling channel ($\gamma_{\text{opt}}$ and $\bar{n}_{\text{opt}}$) and one heat bath ($\gamma_{1}$ and $\bar{n}_{1}$), and the mode $b_{2}$ is contacted with the heat bath ($\gamma_{2}$ and $\bar{n}_{2}$), as shown in Fig.~\ref{Figmodel}(b). Without loss of generality, we assume that the resonance frequencies of two mechanical resonators are the same, namely $\omega_{1}=\omega_{2}=\omega_{m}$. By a lengthly calculation (see Appendix~\ref{appendixb}), the approximate expressions of the final mean phonon numbers can be obtained as
\begin{figure}[tbp]
\centering
\includegraphics[bb=4 16 394 264, width=0.47 \textwidth]{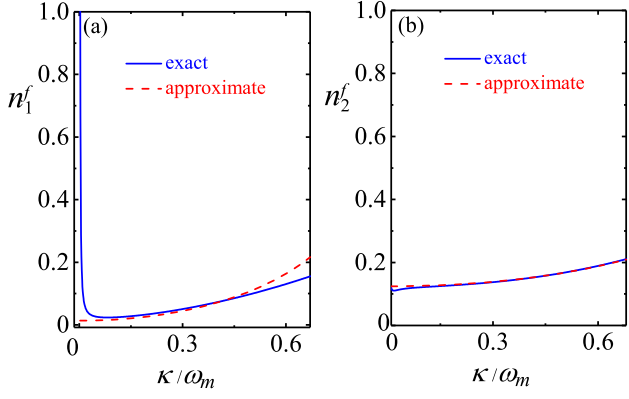}
\caption{(Color online) The final mean phonon numbers (a) $n^{f}_{1}$ and (b) $n^{f}_{2}$ as a function of $\kappa/\omega_{m}$. The exact results are given by Eq.~(\ref{exactcoolresult}) (blue solid curve) and the approximate results obtained by the adiabatic elimination method are given by Eq.~(\ref{coolfina}) (red dashed curve). In addition, we take $\Delta=\omega_{m}$, $\gamma_{1}/\omega_{m}=\gamma_{2}/\omega_{m}=10^{-6}$, and $\eta_{0}/\omega_{1}=0.02$. Other parameters are the same as those given in  Fig.~\ref{detuningdecay}.}\label{comparekappa}
\end{figure}
\begin{subequations}
\label{coolfina}
\begin{align}
n^{f}_{1}\approx&\frac{\gamma_{1}\bar{n}_{1}}{\Gamma_{1}}+\frac{\gamma_{\text{opt}}n_{\text{opt}}+\chi n_{1,\chi }}{\Gamma_{1}-4\chi},\\
n^{f}_{2}\approx&\frac{\gamma_{2}\bar{n}_{2}+\chi n_{2,\chi}}{\chi+\gamma_{2}},
\end{align}
\end{subequations}
with
\begin{subequations}
\begin{align}
n_{\text{opt}}=&\frac{\kappa^{2}}{4(\omega_{m}+\Delta)^{2}},\\
n_{1,\chi}=&\frac{\gamma_{2}\bar{n}_{2}(4\chi+\Gamma_{1})}{(\Gamma_{1}+\gamma_{2})(\chi+\gamma_{2})},\\
n_{2,\chi}=&\frac{\gamma_{1}\bar{n}_{1}+\gamma_{2}\bar{n}_{2}+\gamma_{\text{opt}}n_{\text{opt}}}{\Gamma_{1}+\gamma_{2}},
\end{align}
\end{subequations}
where $\Gamma_{1}=\gamma_{1}+\gamma_{\textrm{opt}}$. We also introduce the effective decay rates $\gamma_{\textrm{opt}}=4\vert\tilde{G}\vert^{2}/\kappa$ and $\chi=4\eta_{0}^{2}/(\gamma_{1}+\gamma_{\text{opt}})$ corresponding to the optomechanical channel and the mechanical coupling channel, respectively. The parameter relations in this case are
\begin{eqnarray}
\omega_{1,2}\gg\kappa\gg\tilde{G}\gg\{\Gamma_{1},\gamma_{\text{opt}}\}\gg\gamma_{1,2}.
\end{eqnarray}
In the optimal-detuning case $\Delta=\omega_{m}$, the corresponding cooling limits $n_{1}^{\lim}$ and $n_{2}^{\lim}$ can be obtained with $n_{\text{opt}}=\kappa^{2}/(16\omega_{m}^{2})$.

\begin{figure}[tbp]
\centering
\includegraphics[bb=6 11 399 260, width=0.47 \textwidth]{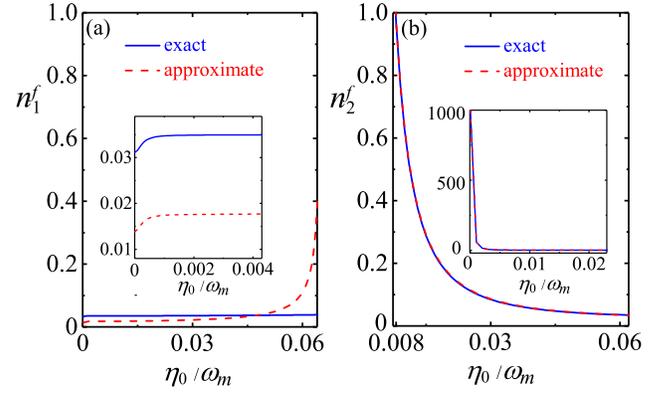}
\caption{(Color online) The final mean phonon numbers (a) $n^{f}_{1}$ and (b) $n^{f}_{2}$ as a function of $\eta_{0}/\omega_{m}$. The exact results (blue solid curve) and the approximate results (red dashed curve) are given by Eq.~(\ref{exactcoolresult}) and Eq.~(\ref{coolfina}), respectively. The insets are zoom-in plots of the phonon numbers in a narrower range of $\eta_{0}/\omega_{m}$. We take $\Delta=\omega_{m}$, $\kappa/\omega_{1}=0.2$, and $P_{L}=70$ mW. Other parameters are the same as those given in Fig.~\ref{comparekappa}.}\label{compareeta}
\end{figure}
To evaluate the approximate cooling results, we compare the approximate results given in Eq.~(\ref{coolfina}) with the exact results given in Eq.~(\ref{exactcoolresult}). In Figs.~\ref{comparekappa} and~\ref{compareeta}, we plot the final mean phonon numbers $n^{f}_{1}$ and $n^{f}_{2}$ as a function of $\kappa$ and $\eta_{0}$ when the optimal effective detuning $\Delta=\omega_{m}$. It shows that the approximate and exact mean phonon numbers coincide well with each other in $\kappa\approx0.1\omega_{m}\sim0.5\omega_{m}$ and $\eta_{0}\approx0\sim0.05\omega_{m}$. Figure~\ref{comparekappa}(a) shows that the difference between the approximate result and the exact result increases when $\kappa<0.1\omega_{m}$. This is because the adiabatic elimination procedure only works under the condition $\kappa\gg\tilde{G}$. In Fig.~\ref{compareeta}(a), we see that the two results do not matched well for a large $\eta_{0}$ (for example $\eta_{0}/\omega_{m}>0.05$ in our simulations). This phenomenon can be explained based on the parameter requirement of the stability in the approximate analyses after the elimination of the cavity field. As shown in Eq.~(\ref{mattixM}), we can see that, to ensure the stability of the equations of motion, the real part of the eigenvalues of the coefficient matrix $\mathbf{M}$ should be positive~\cite{Gradstein2014book}. In the case of $\Omega_{1}\approx\omega_{2}$ and $\gamma_{1}=\gamma_{2}$, the parameter condition is reduced to $\gamma_{\text{opt}}>4\chi$. Corresponding to Fig.~\ref{compareeta}(b), when $\eta_{0}/\omega_{m}>0.05$, the stability condition $\gamma_{\text{opt}}>4\chi$ of the equations of motion in the approximate analyses is violated.

The key physical mechanism in this cooling scheme is that the effective optical vacuum bath successively extracts the excitation energy from the two mechanical modes through the optomechanical cooling channel and the mechanical coupling channel. This physical picture can also be seen from the parameter relation $\gamma_{\text{opt}}>4\chi\gg\gamma_{1,2}$, which indicates that the rate of cooling channel should be much larger than the thermalization channel. The physical picture can also be seen by analyzing the following special cases. When we turn off the mechanical coupling channel, i.e., $\eta=0$, then the first mechanical resonator will be cooled in the same manner as the typical optomechanical sideband cooling scheme~\cite{Wilson-Rae2007PRL,Marquardt2007PRL}, and the second resonator will be thermalized to a thermal equilibrium state at the same temperature as its bath.

\section{Cooling of a chain of coupled mechanical resonators\label{sec5}}

In this section, we extend the optomechanical cooling means to the cooling of a coupled-mechanical-resonator chain. Concretely, we consider an optomechanical cavity coupled to an array of $N$ mechanical resonators connected in series. The nearest neighboring mechanical resonators are coupled to each other through ``position-position" coupling. Without loss of generality, we assume that all the mechanical resonators are identical, having the same frequency, decay rate, and thermal occupation number. Meanwhile, the couplings between the mechanical resonators are much smaller than the mechanical frequency and hence the rotating-wave approximation is justified. Similarly, we consider the strong driving case of the cavity and then perform the linearization procedure to the system. In this case, the Hamiltonian of the system can be written in a frame rotating at the driving frequency as
\begin{eqnarray}
H_{I} &=&\Delta a^{\dagger}a+\omega_{m}\sum^{N}_{j=1}b_{j}^{\dagger}b_{j}-(Ga^{\dagger}b_{1}+G^{*}b_{1}^{\dagger}a)\notag \\
&&-\sum^{N-1}_{j=1}\eta_{0}(b_{j}^{\dagger}b_{j+1}+b_{j+1}^{\dagger}b_{j})+\Omega(a^{\dagger}+a),\label{chainH}
\end{eqnarray}
where $a$ ($a^{\dagger}$) and $b_{j=1-N}=(q_{j}+ip_{j})/\sqrt{2}$ [$b^{\dagger}_{j}=(q_{j}-ip_{j})/\sqrt{2}$] are the annihilation (creation) operators of the cavity mode and the $j$th resonator. The parameter $\Delta$ is the driving detuning after the linearization of the optomechanical coupling, $G$ is the strength of the linearized optomechanical coupling, and $\omega_{m}$ and $\eta_{0}$ are the frequency of these resonators and the coupling strength between the neighboring mechanical resonators, respectively. To include the dissipations, we assume that the cavity is coupled to a vacuum bath and the mechanical resonators are coupled to independent heat baths at the same temperatures. Then the evolution of the system can be governed by the quantum master equation
\begin{eqnarray}
\dot{\rho} &=&i[\rho,H_{I}]+\frac{\kappa}{2}(2a\rho a^{\dagger }-a^{\dagger }a\rho-\rho a^{\dagger }a) \nonumber \\
&&+\frac{\gamma_{m}}{2}(\bar{n}_{m}+1)\sum_{j=1}^{N}(2b_{j}\rho b_{j}^{\dagger }-b_{j}^{\dagger }b_{j}\rho -\rho b_{j}^{\dagger}b_{j})  \nonumber \\
&&+\frac{\gamma_{m}\bar{n}_{m}}{2}\sum_{j=1}^{N}(2b_{j}^{\dagger }\rho b_{j}-b_{j}b_{j}^{\dagger }\rho -\rho b_{j}b_{j}^{\dagger }),\label{chainro}
\end{eqnarray}
where $\rho$ is the density matrix of the coupled cavity-resonator system, $\bar{n}_{m}$ is thermal phonon number of the heat baths of these mechanical resonators, $\kappa$ and $\gamma_{m}$ are the decay rates of the cavity mode and the mechanical resonators, respectively.

\begin{figure}[tbp]
\centering
\includegraphics[bb=22 23 400 259, width=0.47 \textwidth]{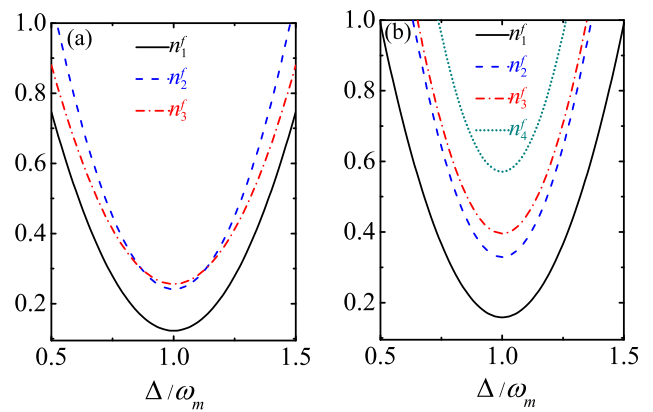}
\caption{(Color online) The final mean phonon numbers in the mechanical resonators as a function of the effective driving detuning $\Delta$ when (a) $N=3$ and (b) $N=4$. Other parameters are given by $G/\omega_{m}=0.2$, $\eta_{0}/\omega_{m}=0.1$, $\kappa/\omega_{m}=0.3$, $\gamma_{m}/\omega_{m}=10^{-5}$, and $\bar{n}=1000$.}\label{coolingchain}
\end{figure}
To evaluate the cooling efficiency, we solve the steady-state solution of quantum master equation~(\ref{chainro}) and calculate the average occupation numbers in the cavity and these mechanical resonators. As examples, we consider the cases of three and four mechanical resonators (i.e., $N=3,4$) in our simulations. In Fig.~\ref{coolingchain}, we plot the final mean phonon numbers in these mechanical resonators as a function of the effective driving detuning $\Delta$ for the cases of (a) $N=3$ and (b) $N=4$. We see that the ground state cooling is achievable and the final phonon numbers successively increases from $n^{f}_{1}$ to $n^{f}_{N}$ at the optimal effective detuning $\Delta=\omega_{m}$. This means that the closer to the optomechanical cavity the resonator is, the smaller the final phonon number in this resonator is. The physical reason for this phenomenon is that the system is a cascade system and the vacuum bath of the optomechanical cavity provides the cooling reservoir to extract the thermal excitations in these mechanical resonators, which are thermally excited by their heat baths. After the linearization, the system is reduced to an array of coupled bosonic modes. Then the vacuum bath provides the cooling channel of the cavity, and the cavity provides the cooling channel of the first mechanical resonator. Successively, the former resonator provides the cooling channel for the next resonator. In this way, the thermal occupations can be extracted to the vacuum bath and then the system approaches to a nonequilibrium steady state. As a result, the cooling efficiency is higher for a mechanical oscillator which is closer to the cavity.

\section{Conclusion\label{sec6}}

In conclusion, we have proposed a scheme to realize the ground-state cooling of coupled mechanical resonators in a three-mode optomechanical system where an optomechanical cavity is coupled to another mechanical resonator. By the linearization, the system is reduced to a cascade-type three-mode coupled system, then the thermal excitations in the mechanical resonators can be extracted to the vacuum bath of the cavity, and the system can be cooled by the optomechanical coupling channel. We found that the coupled mechanical resonators can be simultaneously cooled to their ground states when the system works in the resolved-sideband regime and under a proper driving frequency. In the large-decay limit, we derived analytical expressions of the cooling limits by adiabatically eliminating the cavity field. We also extend the optomechanical method to the cooling of a chain of coupled mechanical resonators. The numerical results show that the ground state cooling is achievable in this system.

\begin{acknowledgments}
D.-G.L. thanks Mr. Wang-Jun Lu for valuable discussions. J.-Q.L. thanks Profs. Yong Li, Xin-You L\"{u}, and Yong-Chun Liu for helpful discussions, and thanks Mr. Da Xu for reading the manuscript and helpful suggestions.
J.-Q.L. thanks Dr. Claudiu Genes for pointing out the integral formula used in Ref.~\cite{Genes2008PRA}. This work is supported in part by National Natural Science Foundation of China
under Grants No.~11774087, No.~11505055, and  No.~11654003, and Hunan Provincial Natural Science Foundation of China under Grant No.~2017JJ1021.
\end{acknowledgments}

\begin{widetext}
\appendix
\section{Calculation of the final mean phonon numbers\label{appendixa}}

In this appendix, we present the detailed calculations of the final mean phonon numbers in the two mechanical resonators. As shown in Sec.~\ref{secabpn}, the exact results of the final mean phonon numbers in the two mechanical resonators can be obtained by calculating the integral in Eq.~(\ref{specintegral}) for the position and momentum fluctuation spectra. Below, we consider the high-temperature limit case $k_{B}T_{l}\gg\hbar\omega_{1,2}$, then it is safe to perform the approximation
\begin{eqnarray}
\gamma_{l}\frac{\omega}{\omega_{l}}\coth\left(\frac{\hbar\omega}{2k_{B}T_{l}}\right)
\approx\gamma_{l}(2\bar{n}_{l}+1),\hspace{1 cm}l=1,2.
\end{eqnarray}
In this case, the integral kernels in Eq.~(\ref{specintegral}) take the form as $g_{n}(\omega)/[h_{n}(\omega)h_{n}(-\omega)]$. This kind of integral can be calculated exactly by the following formula~\cite{Gradstein2014book}:
\begin{eqnarray}
\int_{-\infty}^{\infty}\frac{g_{n}(\omega)}{h_{n}(\omega) h_{n}(-\omega)}d\omega=\frac{i\pi}{a_{0}}\frac{M_{n}}{\Delta_{n}},\label{intgralformula}
\end{eqnarray}
where the functions $g_{n}(\omega)$ and $h_{n}(\omega)$ in the integral kernels take the form as
\begin{eqnarray}
g_{n}(\omega) =b_{0}\omega^{2n-2}+b_{1}\omega^{2n-4}\cdots+b_{n-1}, \hspace{1 cm}h_{n}(\omega) =a_{0}\omega^{n}+a_{1}\omega ^{n-1}\cdots+a_{n},
\end{eqnarray}
with $b_{0,1,2...}$ and $a_{0,1,2,...}$ being the coefficients.
The variables $\Delta_{n}$ and $M_{n}$ in Eq.~(\ref{intgralformula}) are defined by
\begin{eqnarray}
\Delta _{n}=\left\vert\begin{array}{ccccc}
a_{1} & a_{3} & a_{5} & \cdots  & 0 \\
a_{0} & a_{2} & a_{4} &  & 0 \\
0 & a_{1} & a_{3} &  & 0 \\
\vdots  &  &  & \ddots  &  \\
0 & 0 & 0 &  & a_{n}
\end{array}\right\vert,
\hspace{1 cm}
M_{n}=\left\vert\begin{array}{ccccc}
b_{0} & b_{1} & b_{2} & \cdots  & b_{n-1} \\
a_{0} & a_{2} & a_{4} &  & 0 \\
0 & a_{1} & a_{3} &  & 0 \\
\vdots  &  &  & \ddots  &  \\
0 & 0 & 0 &  & a_{n}
\end{array}\right\vert,
\end{eqnarray}
where $|\cdot|$ stands for the determinant. By using the above formula, the integrals in Eq.~(\ref{specintegral}) can be calculated exactly and then the final mean phonon numbers in the two mechanical resonators can be obtained as ($n=6$ for our three mode system)
\begin{eqnarray}
\label{exactcoolresult}
n_{1}^{f} =\frac{1}{2}\left(\frac{iD^{(1)}_{6}}{2\Delta_{6}}+\frac{iM^{(1)}_{6}}{2\Delta_{6}}-1\right),\hspace{1 cm}
n_{2}^{f} =\frac{1}{2}\left(\frac{iD^{(2)}_{6}}{2\Delta_{6}}+\frac{iM^{(2)}_{6}}{2\Delta_{6}}-1\right).
\end{eqnarray}
Here, we introduce the variables
\begin{eqnarray}
\Delta_{6}&=&a_{5}\{a_{4}(-a_{1}a_{2}a_{3}+a_{3}^{2}+a_{1}^{2}a_{4})+[-a_{2}a_{3}+a_{1}(a_{2}^{2}-2a_{4})]a_{5}+a_{5}^{2}\}\nonumber\\
&&-[a_{3}^{3}-a_{1}a_{3}(a_{2}a_{3}+3a_{5})+a_{1}^{2}(a_{3}a_{4}+2a_{2}a_{5})]a_{6}+a_{1}^{3}a_{6}^{2},\label{Delta6def}
\end{eqnarray}
\begin{eqnarray}
D^{(s=1,2)}_{6}&=&[-a_{3}a_{4}a_{5}+a_{3}^{2}a_{6}+a_{5}(a_{2}a_{5}-a_{1}a_{6})] b^{(s)}_{1}+(a_{1}a_{4}a_{5}-a_{5}^{2}-a_{1}a_{3}a_{6})b^{(s)}_{2}\nonumber \\
&&+(-a_{1}a_{2}a_{5}+a_{3}a_{5}+a_{1}^{2}a_{6}) b^{(s)}_{3} +[ -a_{3}^{2}-a_{1}^{2}a_{4}+a_{1}( a_{2}a_{3}+a_{5})] b^{(s)}_{4} \nonumber \\
&&+\frac{1}{a_{6}}[a_{3}^{2}a_{4}-a_{2}a_{3}a_{5}+a_{5}^{2}+a_{1}^{2}(a_{4}^{2}-a_{2}a_{6})+a_{1}(-a_{2}a_{3}a_{4}+a_{2}^{2}a_{5} \nonumber \\
&&-2a_{4}a_{5}+a_{3}a_{6})]b^{(s)}_{5},
\end{eqnarray}
and
\begin{eqnarray}
M^{(s=1,2)}_{6}&=&\frac{1}{\omega_{s}^{2}}
\{-[a_{5}(-a_{2}a_{3}a_{4}+a_{2}^{2}a_{5}+a_{4}(a_{1}a_{4}-a_{0}a_{5}))+(-a_{1}a_{3}a_{4}+a_{0}a_{3}a_{5}+a_{2}(a_{3}^{2}-2a_{1}a_{5}))a_{6}+a_{1}^{2}a_{6}^{2}]b^{(s)}_{1}\nonumber\\
&&+[-a_{3}a_{4}a_{5}+a_{3}^{2}a_{6}+a_{5}(a_{2}a_{5}-a_{1}a_{6}) ]b^{(s)}_{2}+(a_{1}a_{4}a_{5}-a_{5}^{2}-a_{1}a_{3}a_{6})b^{(s)}_{3}+(-a_{1}a_{2}a_{5}+a_{3}a_{5}+a_{1}^{2}a_{6}) b^{(s)}_{4}  \nonumber \\
&&+[ -a_{3}^{2}-a_{1}^{2}a_{4}+a_{1}(a_{2}a_{3}+a_{5})] b^{(s)}_{5}\},
\end{eqnarray}
where the coefficients in our three-mode system are defined by
\begin{eqnarray}
a_{0} &=&1, \nonumber \\
a_{1} &=&-i(2\kappa +\gamma_{1}+\gamma _{2}),  \nonumber \\
a_{2} &=&-[\kappa ^{2}+\gamma _{1}\gamma _{2}+\omega _{1}^{2}+\omega_{2}^{2}+\Delta ^{2}+2\kappa(\gamma_{1}+\gamma_{2})],\nonumber \\
a_{3} &=&i[(\kappa^{2}+\Delta^{2})(\gamma_{1}+\gamma_{2})+2\kappa(\gamma_{1}\gamma_{2}+\omega_{1}^{2}+\omega_{2}^{2})+\gamma _{2}\omega _{1}^{2}+\gamma _{1}\omega _{2}^{2}],\nonumber \\
a_{4} &=&(\kappa ^{2}+\Delta^{2})(\gamma_{1}\gamma_{2}+\omega_{1}^{2}+\omega_{2}^{2}) +2\kappa(\gamma_{2}\omega_{1}^{2}+\gamma_{1}\omega_{2}^{2})+\omega_{1}\omega_{2}(\omega _{1}\omega_{2}-4\eta_{0} ^{2})-2\omega _{1}\vert G\vert^{2}\Delta, \nonumber \\
a_{5} &=&-i\left\{\kappa ^{2}(\gamma _{2}\omega_{1}^{2}+\gamma _{1}\omega_{2}^{2}) +2\kappa \omega _{1}\omega _{2}(\omega_{1}\omega_{2}-4\eta_{0} ^{2}) \right.\left.+\Delta[ \gamma_{1}\omega_{2}^{2}\Delta +\gamma _{2}\omega_{1}(-2\vert G\vert ^{2}+\omega _{1}\Delta)]\right\}, \nonumber \\
a_{6} &=&\omega _{1}\omega _{2}\left\{\Delta [2\omega _{2}\vert G\vert ^{2}-\omega _{1}\omega _{2}\Delta +4\eta_{0} ^{2}\Delta]\right.\left.+\kappa ^{2}( -\omega _{1}\omega _{2}+4\eta_{0} ^{2}) \right\},
\end{eqnarray}
\begin{eqnarray}
b_{0}^{(1)} &=&0,  \nonumber\\
b_{1}^{(1)} &=&(1+2\bar{n}_{1})\gamma _{1}\omega _{1}^{2},  \nonumber\\
b_{2}^{(1)} &=&b_{1}^{(1)}[2\kappa ^{2}+\gamma _{2}^{2}-2(\omega_{2}^{2}+\Delta ^{2})]+2\omega _{1}^{2}\kappa |G|^{2},  \nonumber\\
b_{3}^{(1)} &=&b_{1}^{(1)}[\kappa ^{4}+2\kappa ^{2}(\gamma _{2}^{2}-2\omega_{2}^{2}+\Delta ^{2})+\omega _{2}^{4}-2\gamma _{2}^{2}\Delta ^{2}+\Delta ^{4}+4\omega_{2}^{2}\Delta ^{2}]\nonumber\\
&&+4b_{1}^{(2)}\eta _{0}^{2}\omega _{1}^{2}+2\omega _{1}^{2}|G|^{2}\kappa\lbrack \kappa ^{2}+\gamma _{2}^{2}-2\omega _{2}^{2}+\Delta ^{2}],  \nonumber\\
b_{4}^{(1)} &=&b_{1}^{(1)}[2\omega _{2}^{4}(\kappa ^{2}-\Delta ^{2})+(\gamma_{2}^{2}-2\omega _{2}^{2})(\kappa ^{2}+\Delta ^{2})^{2}]+8b_{1}^{(2)}\eta _{0}^{2}\omega
_{1}^{2}(\kappa ^{2}-\Delta ^{2})\nonumber\\
&&+2\kappa \omega _{1}^{2}|G|^{2}[\omega _{2}^{4}+(\gamma_{2}^{2}-2\omega _{2}^{2})(\kappa ^{2}+\Delta ^{2})],  \nonumber\\
b_{5}^{(1)} &=&(b_{1}^{(1)}\omega _{2}^{4}+4b_{1}^{(2)}\eta _{0}^{2}\omega
_{1}^{2})(\kappa ^{2}+\Delta ^{2})^{2}+2\kappa \omega _{1}^{2}\omega_{2}^{4}(\kappa ^{2}+\Delta ^{2})|G|^{2},
\end{eqnarray}
and
\begin{eqnarray}
b_{0}^{(2)} &=&0, \nonumber\\
b_{1}^{(2)} &=&(1+2\bar{n}_{2})\gamma _{2}\omega _{2}^{2},  \nonumber\\
b_{2}^{(2)} &=&b_{1}^{(2)}[2\kappa ^{2}+\gamma _{1}^{2}-2(\omega_{1}^{2}+\Delta ^{2})], \nonumber\\
b_{3}^{(2)} &=&b_{1}^{(2)}[\kappa ^{4}+2\kappa ^{2}(\gamma _{1}^{2}-2\omega_{1}^{2}+\Delta ^{2})+\Delta ^{2}(\Delta ^{2}-2\gamma _{1}^{2}+4\omega _{1}^{2})+\omega _{1}^{4}-4|G|^{2}\omega _{1}\Delta ]+4b_{1}^{(1)}\omega _{2}^{2}\eta _{0}^{2}, \nonumber\\
b_{4}^{(2)} &=&b_{1}^{(2)}[4|G|^{2}\Delta \omega _{1}(2\kappa \gamma_{1}+\kappa ^{2}+\omega _{1}^{2}+\Delta ^{2})+(\gamma _{1}^{2}-2\omega _{1}^{2})(\kappa ^{2}+\Delta ^{2})^{2}+2\omega_{1}^{4}(\kappa ^{2}-\Delta^{2})]\nonumber\\
&&+8b_{1}^{(1)}\omega _{2}^{2}\eta _{0}^{2}(\kappa ^{2}-\Delta^{2})+8|G|^{2}\kappa (\omega _{1}\omega _{2}\eta _{0})^{2},  \nonumber\\
b_{5}^{(2)} &=&b_{1}^{(2)}\omega _{1}^{2}\{\kappa ^{4}\omega_{1}^{2}+(-2|G|^{2}+\omega _{1}\Delta )[-2|G|^{2}\Delta ^{2}+(\Delta ^{2}+2\kappa ^{2})\omega _{1}\Delta ]\}+4b_{1}^{(1)}\omega_{2}^{2}\eta _{0}^{2}(\kappa ^{2}+\Delta^{2})^{2}\nonumber\\
&&+8|G|^{2}\kappa (\omega _{1}\omega _{2}\eta_{0})^{2}(\kappa ^{2}+\Delta ^{2}).
\end{eqnarray}
Note that the results given by Eq.~(\ref{exactcoolresult}) are exact but complicated. In the large-decay regime, the cavity field can be adiabatically eliminated and we can then obtain analytical and concise expressions of the cooling limits.

\section{Derivation of Eqs.~(\ref{coolfina})\label{appendixb}}

In this appendix, we show a detailed derivation of the cooling limits which are obtained by adiabatically eliminating the cavity field in the large-decay regime. For calculation convenience, we introduce the annihilation and creation operators of the mechanical modes as
\begin{eqnarray}
b_{l=1,2}=(q_{l}+ip_{l})/\sqrt{2},\hspace{1 cm} b^{\dagger}_{l=1,2}=(q_{l}-ip_{l})/\sqrt{2}.
\end{eqnarray}
The Hamiltonian~(\ref{Hamlt2dimless}) can then be expressed as
\begin{eqnarray}
H_{I}&=&\Delta_{c}a^{\dagger}a+\sum_{l=1,2}\omega_{l}b_{l}^{\dagger}b_{l}-\eta_{0}(b_{1}^{\dagger}+b_{1})(b_{2}^{\dagger}+b_{2})-\frac{1}{\sqrt{2}}\lambda _{0}a^{\dagger}a(b_{1}^{\dagger}+b_{1})+\Omega(a^{\dagger}+a),
\end{eqnarray}
where $\Delta_{c}=\omega_{c}-\omega_{L}$ denotes the detuning between the cavity frequency and the driving frequency. By performing the linearization, we write the operators of the system as a summation of their steady-state values and fluctuations: $a\rightarrow\langle a\rangle_{\textrm{ss}}+\delta a$, $b_{1}\rightarrow\langle b_{1}\rangle_{\textrm{ss}}+\delta b_{1}$, and $b_{2}\rightarrow\langle b_{2}\rangle_{\textrm{ss}}+\delta b_{2}$, where $\langle o\rangle_{\textrm{ss}}$ represents the steady-state value of the operator $o$, and $\delta a$, $\delta b_{1}$, and $\delta b_{2}$ are the corresponding fluctuations. The Langevin equations of these fluctuation operators become
\begin{subequations}
\label{langeeqfluct}
\begin{align}
\delta\dot{a}=&(-\kappa/2-i\Delta)\delta a+i\tilde{G}(\delta b_{1}^{\dagger}+\delta b_{1})+\sqrt{\kappa}a_{\textrm{in}},\\
\delta\dot{b}_{1}=&(-\gamma_{1}/2-i\omega_{1})\delta b_{1}+i\eta_{0}(\delta b_{2}^{\dagger}+\delta b_{2})+(i\tilde{G}^{\star}\delta a+i\tilde{G}\delta a^{\dagger})+\sqrt{\gamma_{1}}b_{\textrm{in},1},\label{langeeqfluctb}\\
\delta\dot{b}_{2}=&(-\gamma_{2}/2-i\omega_{2})\delta b_{2}+i\eta_{0}(\delta b_{1}^{\dagger}+\delta b_{1})+\sqrt{\gamma_{2}}b_{\textrm{in},2},\label{langeeqfluctc}
\end{align}
\end{subequations}
where $\Delta=\Delta_{c}-\lambda_{0}(\langle b_{1}\rangle_{\textrm{ss}}^{\star}+\langle b_{1}\rangle_{\textrm{ss}})/\sqrt{2}$ is the normalized detuning and $\tilde{G}=\lambda _{0}\langle a\rangle_{\textrm{ss}}/\sqrt{2}$ is the strength of the linearized optomechanical coupling.

To obtain the cooling limits of the mechanical modes, we consider the parameter regime $\omega _{1,2}\gg \kappa\gg \tilde{G}\gg \gamma _{1,2}$. In this case, the cavity field can be eliminated adiabatically and then the solution of the operator $\delta a(t)$ at the time scale $t\gg 1/\kappa $ can be obtained as
\begin{eqnarray}
\delta a(t)&\approx&\frac{i \tilde{G}}{\kappa/2+i(\Delta+\omega_{1})}\delta b_{1}^{\dagger}(t)+\frac{i \tilde{G}}{\kappa/2+i(\Delta-\omega_{1})}\delta b_{1}(t)+F_{a,\textrm{in}}(t),\label{dletaaadiaelim}
\end{eqnarray}
where we introduce the noise operator
\begin{eqnarray}
F_{a,\textrm{in}}(t)&=&\sqrt{\kappa }e^{-(\kappa/2+i\Delta)t}\int_{0}^{t}e^{(\kappa/2+i\Delta)
s}a_{\textrm{in}}(s)ds.
\end{eqnarray}

Substitution of Eq.~(\ref{dletaaadiaelim}) into Eqs.~(\ref{langeeqfluctb}) and~(\ref{langeeqfluctc}) leads to the equations of motion
\begin{subequations}
\label{redeqofb1b2}
\begin{align}
\delta \dot{b}_{1}(t)=&-(\Gamma_{1}/2+i\Omega_{1}) \delta b_{1}(t)+i\eta_{0} \delta b_{2}(t)+i\tilde{G}^{\star}F_{a,\textrm{in}}(t)+i\tilde{G}F_{a,\textrm{in}}^{\dagger}(t)+\sqrt{\gamma_{1}}b_{\textrm{in},1}(t),\\
\delta\dot{b}_{2}(t)=&i\eta_{0}\delta b_{1}(t)-(\gamma_{2}/2+i\omega_{2})\delta b_{2}(t)+\sqrt{\gamma_{2}}b_{\textrm{in},2}(t),
\end{align}
\end{subequations}
where $\Gamma_{1}=\gamma_{1}+\gamma_{\textrm{opt}}$ and $\Omega_{1}=\omega_{1}-\omega_{\textrm{opt}}$ with $\gamma_{\textrm{opt}}=4\vert\tilde{G}\vert^{2}/\kappa$ and $\omega_{\textrm{opt}}=\vert\tilde{G}\vert^{2}/(2\omega_{1})$, which denote the decay rate and frequency shift induced by cavity coupling channel, respectively.

The final mean phonon numbers (namely the steady-state expected value of the phonon number operators) can be obtained by solving Eq.~(\ref{redeqofb1b2}). To be concise, we reexpress Eq.~(\ref{redeqofb1b2}) as
\begin{equation}
\mathbf{\dot{v}}(t)=-\mathbf{Mv}(t)+\mathbf{N}(t),\label{eqcomptform}
\end{equation}
where $\mathbf{v}(t)=(\delta b_{1}(t), \delta b_{2}(t))^{T}$, $\mathbf{M}$ and $\mathbf{N}(t)$ are defined by
\begin{equation}
\label{mattixM}
\mathbf{M}=\left(
\begin{array}{cc}
 \Gamma_{1}/2+i\Omega _{1} & -i\eta _{0} \\
-i\eta _{0} &\gamma _{2}/2+i\omega _{2}
\end{array}
\right),
\hspace{1 cm}
\mathbf{N}(t) =\left(\begin{array}{c}
i\tilde{G}^{\star }F_{a,\textrm{in}}(t) +i\tilde{G}F_{a,\textrm{in}}^{\dagger}(t)+\sqrt{\gamma _{1}}b_{\textrm{in},1}(t) \\
\sqrt{\gamma _{2}}b_{\textrm{in},2}(t)
\end{array}
\right).
\end{equation}
The formal solution of Eq.~(\ref{eqcomptform}) can be written as
\begin{equation}
\mathbf{v}(t)=e^{-\mathbf{M}t}\mathbf{v}(0)+e^{-\mathbf{M}t}\int_{t_{0}}^{t}e^{\mathbf{M}s}\mathbf{N}(s)ds.
\end{equation}
The final mean phonon numbers can be obtained by calculating the elements of the variance matrix. By a lengthly calculation, we obtain the approximate analytical expressions for the final mean phonon numbers as

\begin{eqnarray}
n_{1}^{f} &=&\frac{1}{4\vert u\vert ^{2}}\bigg\{\gamma _{1}\bar{n}_{1}\bigg[\frac{\vert u-2(\Gamma _{1}-\gamma _{2})-4i(\Omega _{1}-\omega_{2})\vert ^{2}}{\lambda _{1}+\lambda _{1}^{\star }}+\frac{\vert u+2(\Gamma _{1}-\gamma _{2})+4i(\Omega _{1}-\omega _{2})\vert ^{2}}{%
\lambda _{2}+\lambda _{2}^{\star }}  \notag \\
&&+2\text{Re}\Big[\frac{[u-2(\Gamma _{1}-\gamma _{2})-4i(\Omega _{1}-\omega_{2})][u^{\star }+2(\Gamma _{1}-\gamma _{2})-4i(\Omega _{1}-\omega _{2})]}{\lambda _{1}+\lambda _{2}^{\star }}\Big]\bigg]  \notag \\
&&+\vert \tilde{G}\vert ^{2}\bigg[\frac{(\kappa +\lambda _{1}^{\star}+\lambda _{1})\vert u-2(\Gamma _{1}-\gamma _{2})-4i(\Omega _{1}-\omega_{2})\vert ^{2}}{(\lambda _{1}+\lambda _{1}^{\star })\vert \frac{\kappa }{2}+\lambda _{1}+i\Delta \vert ^{2}}+\frac{(\kappa +\lambda_{2}^{\star }+\lambda _{2})\vert u+2(\Gamma _{1}-\gamma _{2})+4i(\Omega_{1}-\omega _{2})\vert ^{2}}{(\lambda _{2}+\lambda _{2}^{\star})\vert \frac{\kappa }{2}+\lambda _{2}+i\Delta \vert ^{2}}  \notag\\
&&+2\text{Re}\Big[\frac{(\kappa +\lambda _{1}+\lambda _{2}^{\star })[u^{\star}+2(\Gamma _{1}-\gamma _{2})-4i(\Omega _{1}-\omega _{2})][u-2(\Gamma_{1}-\gamma _{2})-4i(\Omega _{1}-\omega _{2})]}{(\lambda _{1}+\lambda_{2}^{\star })(\frac{\kappa }{2}+\lambda _{1}+i\Delta )(\frac{\kappa }{2}+\lambda _{2}^{\star }-i\Delta )}\Big]\bigg]  \notag \\
&&+64\eta _{0}^{2}\gamma _{2}\bar{n}_{2}\frac{(\lambda _{1}+\lambda_{1}^{\star }+\lambda _{2}+\lambda _{2}^{\star })[(\lambda _{1}^{\star}+\lambda _{2})(\lambda _{1}+\lambda _{2}^{\star })+(\lambda _{1}+\lambda_{1}^{\star})(\lambda _{2}+\lambda _{2}^{\star })]}{(\lambda _{1}^{\star}+\lambda _{1})(\lambda _{2}^{\star }+\lambda _{2})(\lambda _{1}^{\star}+\lambda _{2})(\lambda _{1}+\lambda _{2}^{\star })}\bigg\},\label{n1f}
\end{eqnarray}
\begin{eqnarray}
n_{2}^{f} &=&\frac{1}{4\vert u\vert ^{2}}\bigg\{64\eta _{0}^{2}\bigg[\gamma_{1}\bar{n}_{1}\frac{(\lambda _{1}-\lambda _{2})(\lambda _{1}^{\star}-\lambda _{2}^{\star })(\lambda _{1}+\lambda _{1}^{\star }+\lambda_{2}+\lambda_{2}^{\star })}{(\lambda _{1}^{\star }+\lambda _{1})(\lambda
_{2}^{\star }+\lambda _{2})(\lambda _{1}^{\star }+\lambda _{2})(\lambda_{1}+\lambda _{2}^{\star })}+\vert \tilde{G}\vert ^{2}\bigg(\frac{\kappa +\lambda _{1}+\lambda _{1}^{\star }}{(\lambda _{1}+\lambda_{1}^{\star })\vert \frac{\kappa }{2}+\lambda _{1}+i\Delta \vert^{2}}  \notag \\
&&+\frac{\kappa +\lambda _{2}+\lambda _{2}^{\star }}{(\lambda _{2}+\lambda_{2}^{\star })\vert \frac{\kappa }{2}+\lambda _{2}+i\Delta \vert^{2}}-2\text{Re}\Big[\frac{\kappa +\lambda _{1}+\lambda _{2}^{\star }}{(\lambda_{1}+\lambda _{2}^{\star })(\frac{\kappa }{2}+\lambda _{1}+i\Delta )(\frac{%
\kappa }{2}+\lambda _{2}^{\star }-i\Delta )}\Big]\bigg)\bigg]  \notag \\
&&+\gamma _{2}\bar{n}_{2}\bigg[\frac{\vert u+2(\Gamma _{1}-\gamma_{2})+4i(\Omega _{1}-\omega _{2})\vert ^{2}}{\lambda _{1}+\lambda _{1}^{\star}}+\frac{\vert u-2(\Gamma _{1}-\gamma _{2})-4i(\Omega_{1}-\omega_{2})\vert ^{2}}{\lambda _{2}+\lambda _{2}^{\star }}\notag \\
&&+2\text{Re}\Big[\frac{[u^{\star }-2(\Gamma _{1}-\gamma _{2})+4i(\Omega_{1}-\omega _{2})][u+2(\Gamma _{1}-\gamma _{2})+4i(\Omega _{1}-\omega _{2})]}{\lambda _{1}+\lambda _{2}^{\star }}\Big]\bigg]\bigg\},\label{n2f}
\end{eqnarray}
where $\lambda_{1}$ and $\lambda_{2}$ ($\lambda^{\star} _{1}$ and $\lambda^{\star} _{2}$ are  complex conjugate) are the eigenvalues of the coefficient matrix $\mathbf{M}$,
\begin{eqnarray}
\label{eigenvalue}
\lambda _{1,2} &=&\frac{1}{4}\left( \Gamma _{1}+\gamma _{2}\right) +\frac{1}{2}i\left( \Omega _{1}+\omega _{2}\right) \mp \frac{1}{8}u,  \notag \\
u &=&\sqrt{4[(\Gamma_{1}-\gamma_{2})+2i(\Omega_{1}-\omega_{2})]^{2}-64\eta_{0}^{2}}.
\end{eqnarray}
Under the parameter condition $\omega_{1,2}\gg\kappa\gg\tilde{G}\gg\{\Gamma_{1},\gamma_{\text{opt}}\}\gg\gamma _{1,2}$, we have $\omega_{1}\gg\omega_{\textrm{opt}}$. In the case of $\omega_{1}=\omega_{2}=\omega_{m}$, Eqs.~(\ref{n1f}) and~(\ref{n2f}) can then be reduced to the results in Eqs.~(\ref{coolfina}).
\end{widetext}

\end{document}